\documentclass[reprint,superscriptaddress,longbibliography,aps,prb,table]{revtex4-2}

\usepackage[english]{babel}
\usepackage{amsmath,amsthm,amssymb,bbm,amsfonts,graphicx,framed,mathtools,dsfont}
\usepackage{url}
\usepackage{enumerate}
\usepackage{mathrsfs}
\usepackage[colorlinks=true, allcolors=blue]{hyperref}
\usepackage[utf8]{inputenc}
\usepackage{graphicx}
\usepackage{float}
\usepackage{booktabs} 
\usepackage{multirow}
\usepackage{cleveref}

\begin{document}

\title{Structural, electronic, and optical properties of hexagonal GeSn from density functional theory}

\author{Yetkin Pulcu}
\affiliation{Department of Physics, University of Konstanz, 78457 Konstanz, Germany}
\author{J{\'a}nos Koltai}
  \affiliation{Department of Biological Physics, E\"otv\"os Lor\'and University, Budapest, Hungary}
\author{Andor Korm{\'a}nyos}
  \affiliation{Department of Physics of Complex Systems, E\"otv\"os Lor\'and University, Budapest, Hungary}
\author{Guido Burkard}
\affiliation{Department of Physics, University of Konstanz, 78457 Konstanz, Germany}

\date{\today}

\begin{abstract}
Unlike cubic GeSn, which undergoes an indirect-to-direct bandgap transition only above a finite Sn concentration, lonsdaleite (2H) germanium is an intrinsic direct-gap semiconductor. We employ first-principles density functional theory to investigate the structural, electronic, and optical properties of 2H-Ge$_{1-x}$Sn$_{x}$ random alloys in the dilute Sn regime ($x \le 0.10$). Substitutional disorder is modeled using 48-atom special quasirandom structure (SQS) supercells, and the coherent effective band structure is recovered via spectral band unfolding. We show that the semiconducting alloy configurations retain a direct bandgap at the $\Gamma$ point, with a moderate, nearly linear reduction of the bandgap in the dilute regime that shifts the fundamental absorption edge toward the mid-infrared. As the gap approaches zero, its calculated value becomes increasingly sensitive to the atomic configuration and supercell size. Evaluation of the optical transition matrix elements shows that the polarization anisotropy characteristic of pristine 2H-Ge remains observable under dilute Sn alloying. Although alloy disorder relaxes the crystal selection rules, the band edge response remains dominated by light polarized perpendicular to the crystal $c$ axis, whereas the parallel component remains weaker. These results identify dilute hexagonal GeSn as a tunable direct-gap system with a strongly polarization-dependent optical response in the infrared.
\end{abstract}

\maketitle

\section{Introduction}
The integration of efficient light-emitting materials into silicon-based technology remains one of the most persistent challenges in photonics \cite{soref2006past, miller2009device, atabaki2018integrating, Han2022LightSourcesSilicon}. Although silicon (Si) and germanium (Ge) dominate the microelectronics industry, their diamond-cubic phases are inherently limited by indirect bandgaps \cite{cardona2010fundamentals}, which severely inhibit efficient light emission. This limitation has historically necessitated the integration of III-V compound semiconductors for optoelectronic functionality \cite{wirths2015lasing}, a process often complicated by lattice mismatch and thermal budget constraints. There is a demand for group-IV-based materials that can bridge the gap between high-performance electronics and photonics within a monolithic platform. Recent breakthroughs in crystal structure engineering have opened new pathways to overcome these fundamental band-structure limitations. It has been demonstrated that by altering the crystal symmetry from the stable cubic diamond phase (3C) to the metastable hexagonal lonsdaleite phase (2H) \cite{joannopoulos1973electronic, amato2016crystal}, the optoelectronic properties of group-IV semiconductors can be drastically modified. Specifically, theoretical and experimental studies have confirmed that 2H-Ge exhibits a direct band gap \cite{rodl2019accurate, de2014predicted, fadaly2020direct,pulcu2024multiband}, resulting from the folding of the conduction band minimum at the L-point of the cubic Brillouin zone onto the $\Gamma$-point of the hexagonal zone. Previous first-principles calculations further identified pristine 2H-Ge as a pseudodirect semiconductor with a weak and strongly polarization-dependent fundamental transition \cite{rodl2019accurate, Broderick2026}. This symmetry manipulation has sparked intense interest in hexagonal group-IV alloys, with recent reports successfully demonstrating the synthesis of high-quality 2H-Si and 2H-Ge/SiGe nanowires via strain-induced phase transition and template-assisted growth mechanisms \cite{fadaly2020direct, vincent2014novel, hauge2015hexagonal}.

While the hexagonal Si$_{1-x}$Ge$_{x}$ system has been explored extensively, showing a tunable direct bandgap for Ge-rich compositions ($x>0.65$) \cite{cartoixa2017optical, wang2021electronic,fadaly2020direct,borlido2023ensemble}, the emission range is primarily limited to the near-infrared and visible spectrum. To push the operational wavelength further into the mid-infrared (MIR) regime, which is critical for sensing, thermal imaging, and free-space communication, tin (Sn) alloying presents a compelling alternative. 

In the cubic phase, following, e.g., the theoretical work of Ref.~\cite{APL-Pandey-1999}, it was shown that Ge$_{1-x}$Sn$_x$ alloys have a direct gap at moderate Sn concentrations, typically around $7$-$11\%$ for unstrained bulk material
\cite{wirths2015lasing, ghetmiri2014direct, gupta2013achieving, Senaratne2016, moontragoon2007band, Lan2017GeSnPhases, Polak2017GeSnBandStructure}.
 Achieving such compositions are metallurgically challenging due to the low equilibrium solubility of Sn in Ge ($<1\%$) \cite{kouvetakis2006tin, Giunto2024GeSnDefects}. 
 By contrast, 2H-Ge is intrinsically a direct bandgap semiconductor at $\Gamma$, so 2H-GeSn does not need to cross an indirect-to-direct transition in order to provide a direct optical gap.
 Furthermore, recent theoretical studies on cubic GeSn indicate that the significant alloy disorder required to reach the direct-gap regime introduces strong intervalley scattering, which can limit carrier mobility and spin lifetimes \cite{sewell2025unveiling, Sewell2025mobility, Sewell2025spin, Myronov2024doping}. Even in the direct-bandgap cubic phase, the small energy separation between the $\Gamma$ and L valleys often leads to thermal carrier leakage, degrading light emission efficiency at room temperature. 

In contrast, the hexagonal (2H) phase of Ge offers a fundamental advantage: it is a direct-gap semiconductor even in its pure form ($x=0$) \cite{rodl2019accurate, fadaly2020direct}. This implies that hexagonal GeSn alloys 
may not require a compositional threshold to achieve a direct bandgap, thereby avoiding the high Sn concentrations necessary for the indirect-to-direct transition in the cubic phase. Consequently, 2H-GeSn offers a new pathway to efficient MIR emission with potentially lower alloy scattering rates and superior thermal stability compared to its cubic counterpart. 
Therefore the hexagonal phase of GeSn (2H-GeSn) offers a largely unexplored but promising frontier.

The recent theoretical work of  Ref.~\cite{Rao2024} calculated the band structure of hexagonal Ge$_{1-x}$Sn$_{x}$ alloys for low Sn concentrations using a single special quasirandom structure (SQS).  It showed that the band gap can be tuned by increasing the Sn concentration while maintaining a direct band gap at the $\Gamma$ point.  Remarkably,  these calculations suggested that compared with 2H-SiGe\cite{wang2021electronic, Bao2021HexSiGe}, which tunes mainly toward the near-infrared with increasing Si concentration, in 2H-GeSn the absorption edge would shift towards the mid-infrared already at low Sn content. Therefore, 2H-GeSn alloys could achieve a direct band gap of a comparable size as their cubic counterparts
but at lower Sn concentrations. However,  Ref.~\cite{Rao2024} did not calculate  the optical properties of 2H-GeSn, such as optical transition matrix elements or absorption coefficient. This is  important, because pure 2H-Ge is known to be a pseudodirect semiconductor\cite{rodl2019accurate}, meaning that  due to optical selection rules it shows  very low optical activity at the band gap energy.  An important motivation for our work is  to fill this  gap of knowledge about the optical responses of  hexagonal GeSn alloys.

In this work, we present a systematic investigation of the properties of hexagonal Ge$_{1-x}$Sn$_{x}$ using first-principles calculations. Building upon established methodologies used for hex-SiGe, we analyze the impact of Sn incorporation on the crystal lattice and optical activity. Furthermore, we map the transformation of the electronic band structure as the Sn concentration is increased, identifying the composition range for direct-gap behavior using the supercell approach, and evaluating the resulting optical properties, including the dielectric function and absorption coefficients.

We note that the first report on the experimental synthesis of hexagonal GeSn polytypes, including the 2H polytype in the composition range studied here, has appeared very  recently~\cite{hex-GeSn-exp-jacs}. This is clearly an important step to unlock the potential of the  hexagonal phases  of GeSn to become  a  material platform for future optoelectonic devices.

\section{Methods}

We investigate the structural, electronic, and optical properties of hexagonal Ge$_{1-x}$Sn$_{x}$ alloys using first-principles density functional theory (DFT) as implemented in the Vienna Ab initio Simulation Package (VASP) \cite{kresse1996efficient, kresse1999ultrasoft} with the projector augmented wave (PAW) method \cite{blochl1994projector}.

To obtain the electronic structure of 
pure 2H-Ge, we followed the methodology of 
Ref.~\cite{rodl2019accurate}, which validated their calculations against known results on cubic Ge.

To model substitutional disorder, we employed the special quasirandom structure (SQS) approach \cite{zunger1990special}, as implemented in the \texttt{mcsqs} code of the Alloy Theoretic Automated Toolkit (ATAT) \cite{vandewalle2002alloy}.
In the ideal random limit, site occupations are uncorrelated and the corresponding cluster correlations are determined solely by the alloy composition. Because of its finite periodicity, the SQS reproduces these correlations only for the selected local clusters and does not represent disorder on length scales exceeding the supercell.

We performed two sets of calculations. For the first set, we used 48-atom $(3\times2\times2)$ supercells and calculated the band structure. 
Structural relaxation was performed using the Perdew-Burke-Ernzerhof exchange-correlation functional for solids (PBEsol) \cite{perdew2008restoring, borlido2021from, Haas2009}, which was selected for its superior accuracy in predicting the lattice equilibrium properties of hexagonal group-IV polymorphs compared to standard Generalized Gradient Approximation (GGA) functionals. The kinetic energy cutoff for the plane-wave basis set was fixed at 400 eV, and the semi-core Ge 3d and Sn 4d electrons were explicitly treated as valence states to ensure an accurate description of the p-d orbital hybridization. The integration of the Brillouin zone was sampled using a $\Gamma$-centered $4\times6\times3$ k-point grid, and all atomic positions and lattice vectors were fully relaxed until the residual forces on all atoms were below $0.001~\mathrm{eV/\AA}$, corresponding to the force-based convergence criterion $\texttt{EDIFFG}=-0.001~\mathrm{eV/\AA}$. 

Following structural optimization, electronic structures were calculated using the Tran–Blaha modified Becke–Johnson exchange potential \cite{tran2009accurate} combined with the local-density approximation (mBJ-LDA), and spin–orbit coupling (SOC) was included. The mBJ-LDA functional was employed to mitigate the band-gap underestimation of semilocal functionals and to provide a reliable description of the band edges needed for optical characterization. Optical dielectric functions and absorption coefficients were computed within the independent-particle approximation directly from the mBJ-LDA Kohn-Sham eigenvalues and dipole matrix elements \cite{choudhary2018computational, Kim2010}. No empirical scissor correction was applied and  many-body GW and Bethe-Salpeter equation corrections were not included.
We note that the independent-particle optical response at the mBJ-LDA level has been shown to capture the relevant optical trends and polarization anisotropy in related hexagonal alloys \cite{Bao2021HexSiGe}. The very recent work of Ref.~\cite{SiGe-excitons} also found that the absorption edge is not shifted by excitonic effects  in hexagonal SiGe alloys.

In order to check if the extracted band gap values depend on the employed supercell, we performed a second set of calculations with different supercells. The employed supercells
and the supercell specific $k$-point samplings were as follows: $32$ atoms ($2\times2\times2$), $7\times7\times4$; 
and $72$ atoms ($3\times3\times2$), $6\times6\times5$. 
The k-point set for each supercell was chosen in such a way that the k-point density along a given direction remains approximately the same as for the $48$ atom supercell. 
Other details of the calculations are the same as explained above for the $48$ atom supercell. 
The band gaps  were extracted from density of states (DOS) calculations performed with a plane-wave cutoff energy of 400 eV, using $\Gamma$ centered Monkhorst--Pack $k$-point grids and the tetrahedron method with Bl\"ochl corrections. We checked if the band gap given by the DOS corresponds to the band gap at $\Gamma$ point.

The band structure calculations of the 48 atom supercell indicated that for higher concentrations the position of the individual Sn atoms in a given SQS supercell can affect the calculated electronic (e.g., band gap and effective mass) and optical properties, as discussed in Sec.~\ref{sec:electronic-struc} and in the Supplementary~\cite{supplementary}. For this reason  we generated $5$ different SQSs for each supercell and concentration, except in the case of the
computationally most demanding $72$ atom supercell, where 
single SQS results are obtained for Sn concentrations $\le7\%$.
The cluster correlation functions of each SQS most closely reproduce those of an ideal random alloy, but the position of the individual Sn atoms in the supercell can nevertheless be different. We then averaged the properties on these ensembles of $5$ SQSs.

\begin{figure}[t]
    \includegraphics[width=0.5\textwidth]{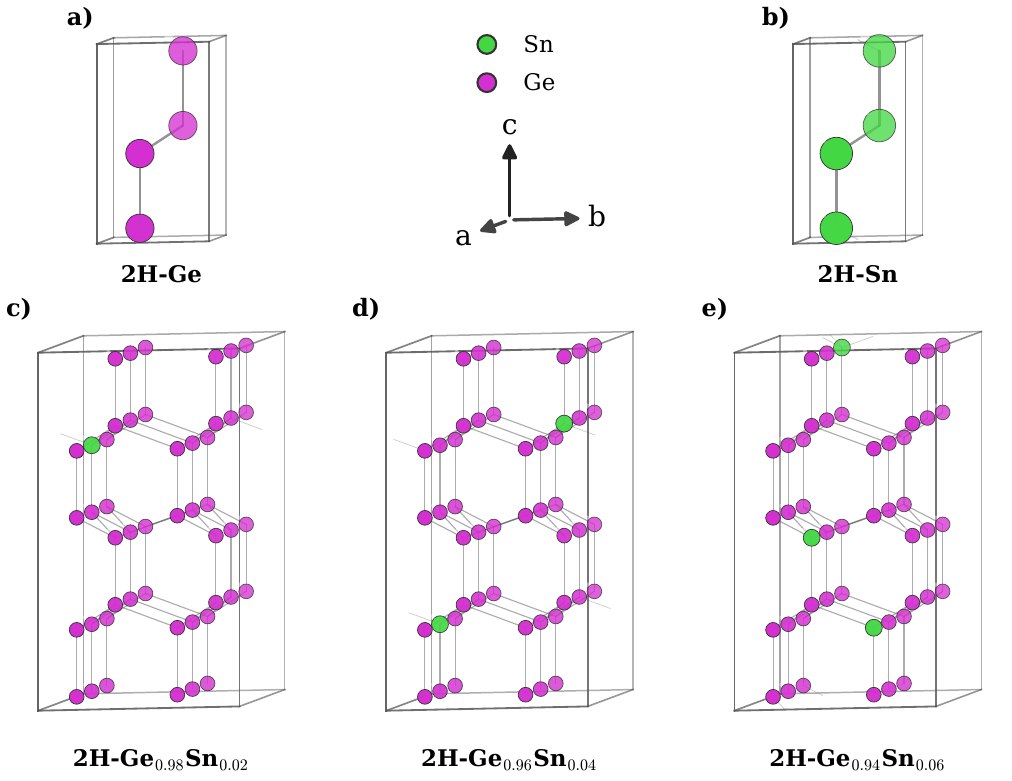} 
   \caption{\label{fig:sige_structures}
   Atomic configurations of hexagonal group-IV systems and random alloys. (a) Primitive cell of 2H-Ge and (b) the hypothetical 2H-Sn phase. (c)--(e) Representative 48-atom $3 \times 2 \times 2$ Special Quasirandom Structure (SQS) supercells for 2H-Ge$_{1-x}$Sn$_{x}$ alloys at concentrations of $x = 0.02$, $0.04$, and $0.06$, respectively. Structures are viewed along the $a$-axis with a tilt.}
\end{figure}

The effective masses reported in Table~\ref{tab:alloy_masses} were obtained from an equal-weight average of the inverse masses, equivalently the band curvatures, according to \[ \left(m_{\mathrm{av}}^*\right)^{-1} = \frac{1}{N_{\mathrm{conf}}} \sum_i \left(m_i^*\right)^{-1}. \]
For the optical quantities in Table~\ref{tab:momentum_matrix}, the squared momentum matrix elements, corresponding to the optical transition strengths, were averaged with equal weight. The absorption spectra in Figs.~\ref{fig:absorption} were also  averaged out. Note, however, that  Figs.~\ref{fig:sige_structures}, \ref{fig:2HGe}, and \ref{fig:bands_evolution} show representative results for a single SQS at a given composition.
\section{Structural Properties}
\label{sec:structural}

The equilibrium crystal structures of the SQS-generated hexagonal $\mathrm{Ge}_{1-x}\mathrm{Sn}_{x}$ alloys were determined by minimizing the total energy with respect to the lattice parameters $a$ and $c$, as well as the internal atomic coordinates. For pure 2H--Ge we obtained lattice constants of $a=$ 3.99\,\AA\ and $c=$ 6.59\,\AA, in excellent agreement with previous PBEsol predictions and experimental data from 2H-Ge nanowires \cite{rodl2019accurate,fadaly2020direct,Bao2021HexSiGe}. We also included a \emph{hypothetical} hexagonal Sn endpoint (2H--Sn) at $x=1$ as a consistent theoretical reference for the alloy series. Although hexagonal Sn is generally treated as a hypothetical phase and is metastable with respect to the $\alpha$- and $\beta$-Sn allotropes, structural relaxations of the pure 2H--Sn cell and intermediate alloys converged while preserving the imposed hexagonal symmetry.\cite{Shao2016HexagonalTinFrameworks, Zeng2015BetaAlphaTin} 

\begin{figure}[t]
    \includegraphics[width=0.5\textwidth]{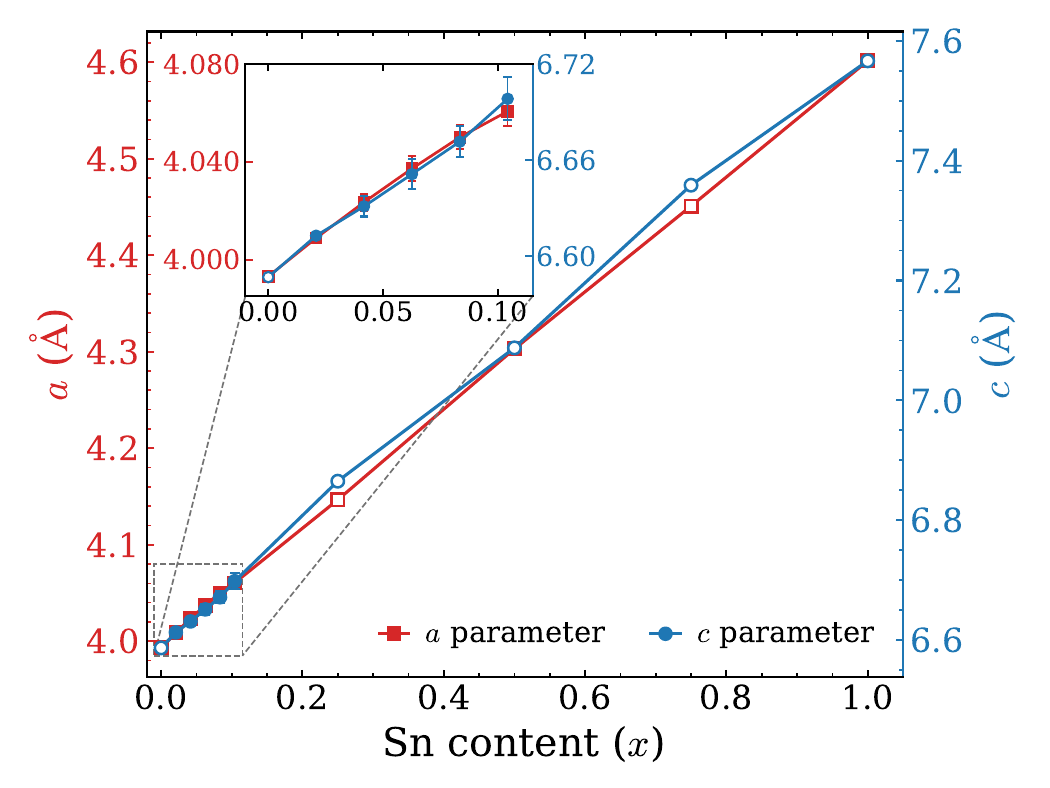} 
      \caption{Calculated lattice parameters $a$ and $c$ of 2H-Ge$_{1-x}$Sn$_x$ as functions of Sn content $x$. For compositions containing one to five Sn atoms in the 48 atom supercell, filled symbols represent averages over the sampled configurations, with error bars showing $\pm$ one sample standard deviation. Open symbols represent the individual calculations for the endpoints and compositions with greater Sn content.}
    \label{fig:lattice_const}
\end{figure}

The evolution of the lattice parameters $a$ and $c$ is shown in Fig.~\ref{fig:lattice_const}. Their deviations from Vegard's Law \cite{vegard1921konstitution} were calculated using
\begin{equation}
P(x)=(1-x)P_{\mathrm{Ge}}+xP_{\mathrm{Sn}}-b_Px(1-x), \quad P=a,c.
\end{equation}
Fitting this to the numerical results yields $b_a=-0.022~\text{\AA}$ and $b_c=-0.114~\text{\AA}$. Under this convention, both negative values correspond to weak upward deviations from linear interpolation \cite{bernard1987electronic}. We also tracked the hexagonal anisotropy via the $c/a$ ratio across the composition range. The ratio remains nearly constant over the entire series, varying only slightly between $c/a\!=\!1.64$ and $c/a\!=\!1.66$. This small, non-monotonic variation (maximal change $\lesssim 1\%$) indicates that Sn incorporation expands the lattice almost isotropically in the hexagonal framework, with only minor anisotropic accommodation along the out-of-plane direction. Small deviations from a smooth trend may arise from local relaxation around Sn atoms in the disordered alloys.

\section{Electronic Structure}
\label{sec:electronic-struc}

Having established the structural stability of the hexagonal alloys, we now turn to their electronic properties.

\subsection{Band Symmetries and Direct Bandgap}

Although the electronic structure of pure 2H-Ge has been discussed in previous works \cite{rodl2019accurate, wang2021electronic,pulcu2024multiband}, we re-evaluated it here to establish a consistent theoretical framework for the hexagonal GeSn alloy system. By employing the identical computational framework using the mBJ-LDA functional and PAW potentials for both pure Ge and for the GeSn SQS supercells, we ensure that the derived trends in bandgap and effective mass evolution with varying Sn concentration are driven by physical alloying effects rather than systematic errors arising from different methodologies.

\begin{figure}
    \includegraphics[width=0.5\textwidth]{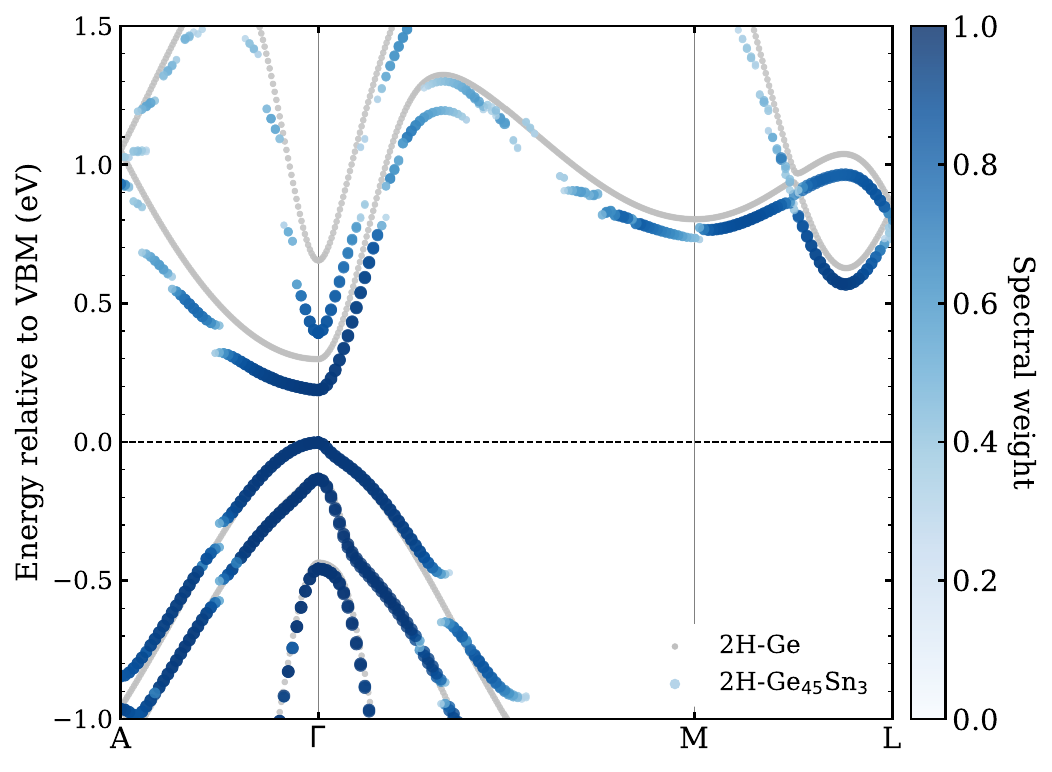} 
      \caption{Comparison of the electronic band structures of pure 2H-Ge (gray lines) and a representative 48-atom Ge$_{45}$Sn$_3$ SQS (blue symbols, $x=3/48=0.0625$), calculated using mBJ-LDA including SOC. The alloy bands were unfolded onto the primitive 2H-Ge Brillouin zone. The energies of each system are referenced to their respective valence band maxima. The marker area and colour intensity scale with the unfolded spectral weight. For clarity, only states with $W_{\mathbf{K},m}(\mathbf{k}_i)\geq0.3$ are shown. The displayed alloy configuration retains a direct fundamental gap at $\Gamma$.}
    \label{fig:2HGe}
\end{figure}

The calculated electronic band structure of pure 2H-Ge is displayed in Fig.~\ref{fig:2HGe}. The direct gap originates from the folding of the cubic \(L\)-point along the \(\langle111\rangle\) direction.
We find a fundamental direct bandgap of \(0.30\)~eV at the \(\Gamma\)-point, in reasonable agreement with the experimental value\cite{fadaly2020direct} of \(\sim0.35\)~eV reported for Ge nanowires and in good agreement with previous DFT results \cite{rodl2019accurate,pulcu2024multiband,wang2021electronic}. We note that an even better  agreement for the band gap was  obtained using  a semi-empirical  methodolgy in the very recent work of Ref.~\cite{Broderick2026}.

The reduced hexagonal symmetry lifts the degeneracy of the valence band maximum (VBM). Unlike in the cubic phase where heavy-hole and light-hole bands are degenerate at $\Gamma$, 2H-Ge exhibits a distinct splitting governed by the interplay of the crystal-field splitting $\Delta_{\mathrm{cf}}$ and the anisotropic spin-orbit coupling $\Delta_{\mathrm{so}}$. Using $\mathbf{k} \cdot \mathbf{p}$ theory, the energy separation between the $\Gamma_{9v}^{+}$ and the crystal-field split-off bands ($\Gamma_{7v}^{\pm}$) is described by the relation~\cite{rodl2019accurate},
\begin{equation}
\begin{split}
    \varepsilon(\Gamma_{9v}^{+}) - \varepsilon(\Gamma_{7v\pm}^{+}) &= \frac{\Delta_{\mathrm{cf}} + \Delta_{\mathrm{so}}^{\parallel}}{2} \\
    &\quad \mp \frac{1}{2}\sqrt{\left(\Delta_{\mathrm{cf}} - \frac{1}{3}\Delta_{\mathrm{so}}^{\parallel}\right)^2 + \frac{8}{9}(\Delta_{\mathrm{so}}^{\perp})^2}.
\end{split}
\end{equation}
Here, $\varepsilon(\Gamma_{9v}^{+})$ corresponds to the energy of the top valence band (VB), while $\varepsilon(\Gamma_{7v\pm}^{+})$ represents second-highest (VB-1) and third-highest (VB-2) valence bands, respectively. Based on this model, our calculations yield a crystal-field splitting of $\Delta_{\mathrm{cf}} = 0.268$~eV, with spin-orbit components of $\Delta_{\mathrm{so}}^{\perp} = 0.278$~eV and $\Delta_{\mathrm{so}}^{\parallel} = 0.286$~eV. As a comparison, we show the corresponding mBJ-LDA and HSE06 results of  Rödl \textit{et al.}~\cite{rodl2019accurate} in Table~\ref{tab:splittings}.   Our mBJ-LDA values agree closely with Ref.~\cite{rodl2019accurate}, whereas the HSE06 values are systematically larger.

\begin{table}[t]
\caption{\label{tab:splittings}
Crystal-field and spin-orbit splittings of 2H-Ge at $\Gamma$.
The values from this work are compared with the HSE06 and mBJ-LDA
results reported by Rödl \textit{et al.}~\cite{rodl2019accurate}. The
direction-averaged spin-orbit splitting is
$\Delta_{\mathrm{so}}^{\mathrm{av}}=
(\Delta_{\mathrm{so}}^{\parallel}
+2\Delta_{\mathrm{so}}^{\perp})/3$.}
\begin{ruledtabular}
\begin{tabular}{lccc}
Quantity
& This work
& Ref.~\cite{rodl2019accurate} mBJ-LDA
& Ref.~\cite{rodl2019accurate} HSE06 \\
\colrule
$\Delta_{\mathrm{cf}}$ (eV)
& 0.268 & 0.270 & 0.288 \\
$\Delta_{\mathrm{so}}^{\parallel}$ (eV)
& 0.286 & 0.282 & 0.329 \\
$\Delta_{\mathrm{so}}^{\perp}$ (eV)
& 0.278 & 0.274 & 0.320 \\
$\Delta_{\mathrm{so}}^{\mathrm{av}}$ (eV)
& 0.281 & 0.277 & 0.323
\end{tabular}
\end{ruledtabular}
\end{table}

To evaluate the carrier transport potential of the hexagonal phase, we calculated the effective masses $m^*$ for the band-edge states along the characteristic high-symmetry directions. The results, summarized in Table~\ref{tab:ge_mass_comparison}, reveal a very large anisotropy in the electronic dispersion arising from the specific $D_{6h}$ crystal symmetry.
For electrons in the conduction band (CB), the effective mass is highly directional. We calculate a very small in-plane effective mass of $m_{e}^{\perp} = 0.088~m_0$ (along $\Gamma \rightarrow M$), which is comparable to the small electron mass in bulk GaAs ($0.067~m_0$) \cite{vurgaftman2001band}. In contrast, the effective mass along the stacking direction ($\Gamma \rightarrow A$) is more than an order of magnitude larger ($m_{e}^{\parallel} = 1.102~m_0$), reflecting the weak band dispersion and strong transport anisotropy along the $c$ axis. This substantial anisotropy can be understood by the selection rules of the $D_{6h}$ point group within the $\mathbf{k} \cdot \mathbf{p}$ theory \cite{chuang1996kp, bir1974symmetry}. Within the  $D_{6h}$ symmetry, the lowest conduction band has no symmetry-allowed momentum matrix element coupling to the valence states along the c-axis at $\Gamma$. Hence, the $\Gamma \rightarrow A$ 
dispersion receives no corresponding to $\mathbf{k} \cdot \mathbf{p}$ contribution and remains relatively flat. For SOC-induced weak optical activity of the fundamental transition instead occurs for in plane polarization, as discussed in \ref{sec:optical}. This weak interaction translates directly into a flat-band dispersion and the large effective mass observed along the c-axis. A similar anisotropy is observed for the valence band holes. The top valence band (VB) exhibits a light-hole character in the in-plane direction ($m_{h}^{\perp} = 0.067~m_0$) but behaves as a heavy hole along the $c$-axis ($m_{h}^{\parallel} = 0.517~m_0$).

We note that the effective mass values in hexagonal polymorphs are sensitive to the non-parabolicity of the bands away from the zone center. To accurately capture the band-edge curvature for the relevant band, our effective masses were derived by fitting the dispersion within a narrow $k$-space range of approximately $|\Delta \mathbf{k}| < 0.06~\text{\AA}^{-1}$ around the $\Gamma$-point. Small deviations from values reported in Ref.~\cite{rodl2019accurate, Rao2024} are attributed to differences in this fitting range and the specific k-point sampling density.

\begin{table*}[t]
\caption{\label{tab:ge_mass_comparison}
Comparison of the calculated electron and hole effective masses of
2H-Ge with the HSE06 and mBJ-LDA results of
Ref.~\cite{rodl2019accurate}. All values are given in units of the
free electron mass $m_0$. All calculations include spin-orbit coupling.}
\begin{ruledtabular}
\begin{tabular}{lcccccc}
& \multicolumn{2}{c}{This work, mBJ-LDA}
& \multicolumn{2}{c}{Ref.~\cite{rodl2019accurate}, mBJ-LDA}
& \multicolumn{2}{c}{Ref.~\cite{rodl2019accurate}, HSE06} \\
Band (irrep.)
& $\Gamma\rightarrow A$ & $\Gamma\rightarrow M$
& $\Gamma\rightarrow A$ & $\Gamma\rightarrow M$
& $\Gamma\rightarrow A$ & $\Gamma\rightarrow M$ \\
\hline
CB+1 ($\Gamma_{7c}^{-}$)  & 0.039 & 0.043 & 0.042 & 0.052 & 0.033 & 0.038 \\
CB   ($\Gamma_{8c}^{-}$)  & 1.102 & 0.088 & 1.088 & 0.089 & 0.997 & 0.076 \\
VB   ($\Gamma_{9v}^{+}$)  & 0.517 & 0.067 & 0.526 & 0.074 & 0.463 & 0.055 \\
VB-1 ($\Gamma_{7v+}^{+}$) & 0.113 & 0.083 & 0.120 & 0.101 & 0.085 & 0.079 \\
VB-2 ($\Gamma_{7v-}^{+}$) & 0.048 & 0.295 & 0.053 & 0.325 & 0.044 & 0.252 \\
\end{tabular}
\end{ruledtabular}
\end{table*}

In order to benchmark the effective mass extraction employed in this work and other band structure parameters, Table~\ref{tab:ge_mass_comparison} compares our mBJ-LDA results for pristine 2H-Ge with the mBJ-LDA and HSE06 values reported in Ref.~\cite{rodl2019accurate}. 
We note that, as shown in Ref.~\cite{Kim2010}, 
mBJ-LDA tends to overestimate the effective masses.
The present results generally lie between the two previous calculations and are particularly close to the corresponding mBJ-LDA values. The remaining differences can arise from the fitting window and numerical details used to extract the band curvatures. This agreement supports the application of the same procedure to the unfolded alloy band structures.

\begin{table}[htbp]
\caption{\label{tab:alloy_masses} 
Calculated electron and hole effective masses of 2H-Ge and 2H-Ge$_{1-x}$Sn$_x$ along the $\Gamma\rightarrow\mathrm{A}$ and $\Gamma\rightarrow\mathrm{M}$ directions, in units of the free electron mass $m_0$. For the alloys, the reported values are obtained by averaging the inverse masses, equivalently the band curvatures, over the sampled atomic configurations and subsequently inverting the result.}
\begin{ruledtabular}
\begin{tabular}{llcc}
Material & Band (irrep.) & $m^*_{\Gamma \rightarrow \text{A}}$ ($m_0$) & $m^*_{\Gamma \rightarrow \text{M}}$ ($m_0$) \\
\hline
\multirow{5}{*}{2H-Ge} & CB+1 $(\Gamma_{7c}^{-})$ & 0.039 & 0.043 \\
                       & CB $(\Gamma_{8c}^{-})$  & 1.102 & 0.088 \\
                       & VB $(\Gamma_{9v}^{+})$   & 0.517 & 0.067 \\
                       & VB-1 $(\Gamma_{7v^+}^{+})$ & 0.113 & 0.083 \\
                       & VB-2 $(\Gamma_{7v^-}^{+})$ & 0.048 & 0.295 \\
\hline 

\multirow{5}{*}{$\text{Ge}_{0.98}\text{Sn}_{0.02}$} & CB+1 & 0.039 & 0.040 \\
                                                    & CB   & 0.694 & 0.084 \\
                                                    & VB   & 0.520 & 0.064 \\
                                                    & VB-1 & 0.097 & 0.071 \\
                                                    & VB-2 & 0.046 & 0.279 \\
\hline

\multirow{5}{*}{$\text{Ge}_{0.96}\text{Sn}_{0.04}$} & CB+1 & 0.037 & 0.036 \\
                                                    & CB   & 0.601 & 0.087 \\
                                                    & VB   & 0.513 & 0.066 \\
                                                    & VB-1 & 0.090 & 0.061 \\
                                                    & VB-2 & 0.044 & 0.271 \\
                                                    
\hline

\multirow{5}{*}{$\text{Ge}_{0.94}\text{Sn}_{0.06}$} & CB+1 & 0.036 & 0.034 \\
                                                    & CB   & 0.507 & 0.080 \\
                                                    & VB   & 0.507 & 0.059 \\
                                                    & VB-1 & 0.082 & 0.056 \\
                                                    & VB-2 & 0.043 & 0.259 \\

\hline

\multirow{5}{*}{$\text{Ge}_{0.92}\text{Sn}_{0.08}$} & CB+1 & 0.034 & 0.032 \\
                                                    & CB   & 0.342 & 0.073 \\
                                                    & VB   & 0.481 & 0.055 \\
                                                    & VB-1 & 0.072 & 0.051 \\
                                                    & VB-2 & 0.041 & 0.243 \\
\end{tabular}
\end{ruledtabular}
\end{table}

\subsection{Electronic Structure of 2H-Ge$_{1-x}$Sn$_{x}$}

Theoretical modeling of random alloy systems often relies on a thermodynamic ensemble average over all possible atomic configurations within a small unit cell. Combined with the generalized quasi-chemical approximation, this method has been successfully applied to 2H-SiGe using 8-atom clusters \cite{borlido2023ensemble}. Such small cells, however, are fundamentally restricted in their compositional resolution, as an 8-atom cell dictates a minimum non-zero solute concentration of 12.5\%. Because our primary focus is the dilute Sn regime ($x \le 10\%$), where preserving the direct bandgap is most critical for optoelectronics, these small-cluster ensembles are inadequate. Expanding the ensemble method to larger supercells (e.g., 16 or 32 atoms) to capture lower Sn concentrations leads to a combinatorial explosion of symmetry-inequivalent configurations, making first-principles calculations computationally very challenging.

\begin{figure}[t]
    \centering
    \includegraphics[width=\linewidth]{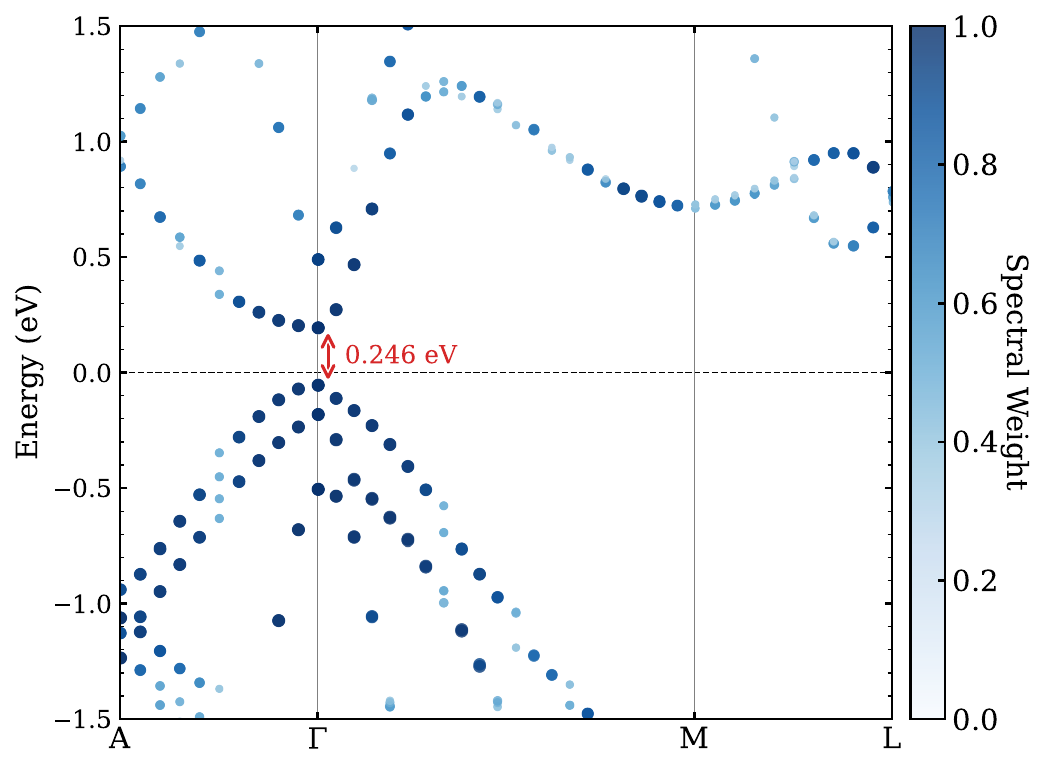}
    \caption{Unfolded electronic band structure of 2H-Ge$_{47}$Sn$_1$ ($x=1/48\approx0.021$), obtained from a representative 48-atom ($3\times2\times2$) SQS using mBJ-LDA including SOC. The marker area and color intensity scale with the unfolded spectral weight. The band edge states retain strong coherent spectral weight at $\Gamma$, and the fundamental gap remains direct, with $E_g=0.246$~eV for the configuration shown.}
    \label{fig:bands_evolution}
\end{figure}

To avoid these limitations, we employ the Special Quasirandom Structure (SQS) approach \cite{zunger1990special}. Importantly, recent theoretical benchmarks on similar
hexagonal SiGe alloys have demonstrated that large SQS supercells yield comparable macroscopic physical properties and thermodynamic stability profiles to full ensemble averages \cite{borlido2023ensemble}. This confirms that the underlying alloy physics remains consistent between the two methods. Employing 48-atom $3\times2\times2$ SQS supercells, we capture the essential local disorder of dilute $\mathrm{Ge}_{1-x}\mathrm{Sn}_x$ alloys within a manageable first-principles framework. Additional calculations, with supercells up to $72$ atoms, were carried out to evaluate finite-size effects and to test the robustness of the band-gap trends against the choice of supercell size. By combining the SQS description with band unfolding, we can track the evolution of the band-edge states under substitutional disorder and assess whether the direct gap at $\Gamma$ is preserved in the dilute alloy limit, as shown in Fig.~\ref{fig:bands_evolution}.

To facilitate direct comparison with the primitive cell band structure of 2H-Ge, we used band unfolding to project the supercell eigenstates onto the primitive 2H-Ge Brillouin zone. We used the method implemented in the \textsc{vaspkit} code \cite{wang2021vaspkit}, which calculates the effective spectral weight $W_{\mathbf{K}}(\mathbf{k}_i)$ \cite{popescu2010effective} for a primitive wave vector $\mathbf{k}_i$ by projecting the supercell eigenstates $|\Psi_{\mathbf{K},m}\rangle$ onto the primitive cell basis states.
Mathematically, the spectral-weight contribution of the $m$-th band of the supercell (at the supercell wave vector $\mathbf{K}$) to the primitive wave vector $\mathbf{k}_i$ is given by,
\begin{equation}
    W_{\mathbf{K},m}(\mathbf{k}_i)
    =
    \sum_{n}
    \left| \langle \phi_{\mathbf{k}_i,n} | \Psi_{\mathbf{K},m} \rangle \right|^2 .
\end{equation}
Here, $\lvert\phi_{\mathbf{k}_i,n}\rangle$ denotes a Bloch state of the primitive cell, and the summation runs over the primitive bands $n$. The projection resolves the supercell eigenstates according to their overlap with the Bloch states of the ideal primitive crystal. In the effective band structure (EBS) shown in Fig.~\ref{fig:bands_evolution}, the marker area and colour intensity scales with the spectral weight $W_{\mathbf{K},m}(\mathbf{k}_i)$. The resulting distribution shows the degree to which the primitive Bloch character is retained and reveals the spectral broadening associated with alloy disorder.

The resulting EBS for the Ge$_{47}$Sn$_{1}$ alloy is presented in Fig.~\ref{fig:bands_evolution}. The projection reveals that the conduction band minimum (CBM) retains significant coherent spectral weight at the $\Gamma$-point, indicating that the Bloch character of the host state is largely preserved despite the symmetry-breaking potential of the Sn substitution. While the unfolded bands away from the band edge show reduced spectral weight and increased broadening, the fundamental band edge states remain well-defined. Crucially, no competing conduction valley is found below the $\Gamma$ valley, confirming that the alloy retains a direct fundamental gap in the dilute regime.

A comparison of the EBS of an Ge$_{45}$Sn$_3$ SQS and the band structure of pure 2H-Ge is shown in 
Fig.~\ref{fig:2HGe}. 
Compared with pure 2H-Ge, 
the Ge$_{45}$Sn$_3$ SQS exhibits a reduced fundamental gap while retaining the overall band edge dispersion and spectral character of the host material near $\Gamma$. This figure also indicates that Sn incorporation narrows the gap 
in the dilute alloy regime.
One can also notice   discontinuities in the dispersion of the EBS, e.g., at half-way along the $\Gamma-A$ line. For lower Sn content this feature is less pronounced, c.f.  Fig.~\ref{fig:bands_evolution}
for Ge$_{47}$Sn$_{1}$. This feature in the EBS is due to the  use of supercells. The bands of pure 2H-Ge would be  degenerate at the supercell BZ boundaries because of  band folding. In the  GeSn alloy calculations these degeneracies are lifted by the presence of the Sn atoms. In the case of the $3\times 2\times 2$ supercells this leads to the apparent discontinuity of the bands e.g., at half-way along the $\Gamma-A$ line. We find that  that EBS of different SQS realizations can display  different level of perturbations even for the same nominal Sn concentration, see the Supplementary~\cite{supplementary}. 
This perturbation of the EBS can also affect the value of the extracted effective masses, especially for the CB along the  $\Gamma-A$ line. This was one of the reasons to generate 
$5$ different SQSs for each Sn concentrations and 
perform averaging of the results.

\begin{figure}[t]
    \includegraphics[width=0.5\textwidth]{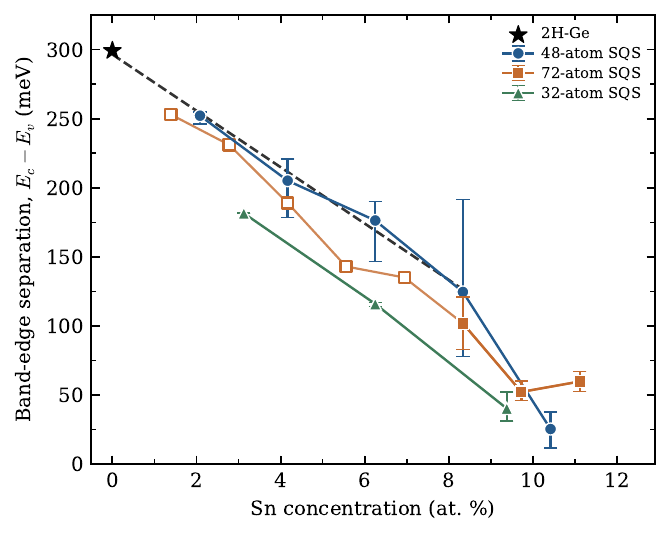}
      \caption{Composition dependence of the 
      band gap in 2H-Ge$_{1-x}$Sn$_x$, calculated using mBJ-LDA including SOC. 
      Filled symbols denote configurational averages obtained 
      from five fully relaxed SQSs,
      with asymmetric bars spanning the complete range of calculated values. 
      Open squares show single-SQS results obtained using the 
      $72$ atom supercells. The star marks pristine 2H-Ge. The dashed line is a linear fit to pristine 2H-Ge and the 48-atom configurational means. 
      Solid lines are guides to the eye.}
        \label{fig:SnVsGe}
\end{figure}

We start the discussion of alloying effects 
with the  dependence of the fundamental band gap at the $\Gamma$ point on 
the Sn concentration. 
In Fig.~\ref{fig:SnVsGe} we show the averaged band gap over 5 SQSs as a function of Sn concentration for three different supercells, 
as well as single SQS results for the largest, 72 atom supercell at lower Sn concentrations.
Where averaged band gap values are given, we also indicate the range of obtained band gap values for each concentration and supercell. As shown by the results of the 
$48$ atom supercell, the introduction of Sn induces a monotonic narrowing of the averaged direct $\Gamma$ point band edge separation. As indicated in Fig.~\ref{fig:SnVsGe}, at approximately 2 at.\% Sn, the mean gap decreases from 0.299 eV in pristine 2H-Ge to $0.252$ eV, with the sampled configurations ranging from $0.246$ to $0.255$ eV. At approximately 8\% Sn, the mean gap is reduced to $0.125$ eV, although the calculated values span a considerably broader range of $0.078-0.191$ eV. A linear fit to pristine 2H-Ge and the 48 atom configurational mean up to approximately 8\% Sn yields a gap reduction of 20.4 meV per \% Sn. The corresponding mean gap energies cover wavelengths from approximately 4.9 $\mu$m to 10.0 $\mu$m, while the unfolded band structures confirm that the semiconducting configurations retain direct band edges at $\Gamma$.

We  performed additional calculations in order to investigate whether band gap narrowing is robust under different supercell shapes.
As shown in Fig.~\ref{fig:SnVsGe}, the same trend can be observed in the  $2\times2\times2$, 32 atom supercell results as for the larger, 48 atom supercell calculations, albeit with smaller gap values. A similar result has been obtained by Ref.~\cite{Rao2024}. However, we note that our band gap values are lower than the corresponding ones in Ref.~\cite{Rao2024} and  for the highest Sn concentration, i.e., the Ge$_{29}$Sn$_3$ supercell, we found that $4$ out of $5$ SQSs yielded an indirect band gap. (For the $48$ and $72$ atom supercells we did not find indirect band gap in the concentration range considered in Fig.~\ref{fig:SnVsGe}.)
The same decreasing 
trend of the band gap for low Sn concentrations can be observed  from single SQS calculations using $3\times3\times2$, $72$ atom supercells. Together with the $32$ and $48$ atom supercell results this shows that  the  decreasing trend of the band gap  does not depend on the supercell shape and size for Sn concentrations $\lesssim 7\%$.

Previous work on the cubic GeSn alloy   showed that local Sn-Sn correlations and short range order (SRO) can affect the calculated band gap, with canonical sampling revealing a tendency toward reduced first-neighbor Sn-Sn pairing \cite{Cao2020SRO}.
Similarly to cubic GeSn,  one may expect that such effects can be  important in hexagonal GeSn alloys for higher Sn concentrations. 
To see how the finite supercell size may influence the band gap in this case, we performed  calculations using the $72$ atom supercells and averaging the band gap over $5$ SQSs. The 
results are also shown in Fig.~\ref{fig:SnVsGe}, along with the results using the $32$ and $48$ atom supercells. Considering all supercell results shown in Fig.~\ref{fig:SnVsGe}, they indicate further gap reduction for Sn concentrations $\lesssim 10\%$.
As the band gap approaches small values in this regime, its precise value may  become  increasingly sensitive to supercell shape and possibly to SRO effects. The exact composition at which gap closure occurs cannot be unambiguously determined from the present supercell calculations and it is beyond the scope of the present work.

Table~\ref{tab:alloy_masses} shows the Sn composition dependence of the in-plane and out of plane effective masses. The effective masses were obtained from the 48 atom $3\times2\times 2$ supercell calculations, by averaging the results from $5$ SQS for each composition.
Upon alloying with Sn, the effective masses retain the strong anisotropy characteristic of the hexagonal phase, as summarized in Table~\ref{tab:alloy_masses}. Over the range $0\leq x\leq0.08$, the in-plane conduction-band mass remains small, varying only between $m_e^\perp=0.088\,m_0$ and $0.073\,m_0$. The corresponding out-of-plane mass exhibits a much stronger composition dependence, decreasing from $m_e^\parallel=1.102\,m_0$ in 2H-Ge to $0.342\,m_0$ at $x=0.08$. Hence, Sn incorporation preserves the light in-plane carrier masses while producing a pronounced reduction of the out-of-plane conduction-band mass.
The top valence band retains the same qualitative anisotropy: its in-plane hole mass varies between $m_h^\perp=0.055\,m_0$ and $0.067\,m_0$, whereas the out-of-plane mass remains considerably larger, between $m_h^\parallel=0.481\,m_0$ and $0.520\,m_0$. 

\section{Optical Properties}
\label{sec:optical}
Having demonstrated the tunability of the electronic band structure, we now turn to the optical response of the hexagonal GeSn alloys. The macroscopic optical properties are intrinsically linked to the microscopic transition probabilities between the valence and conduction bands. For optoelectronic applications, specifically lasing and detection, the magnitude of the dipole transition matrix element is as critical as the directness of the bandgap itself.

\subsection{Momentum Matrix Elements and Selection Rules}

\begin{figure}[t]
    \includegraphics[width=0.5\textwidth]{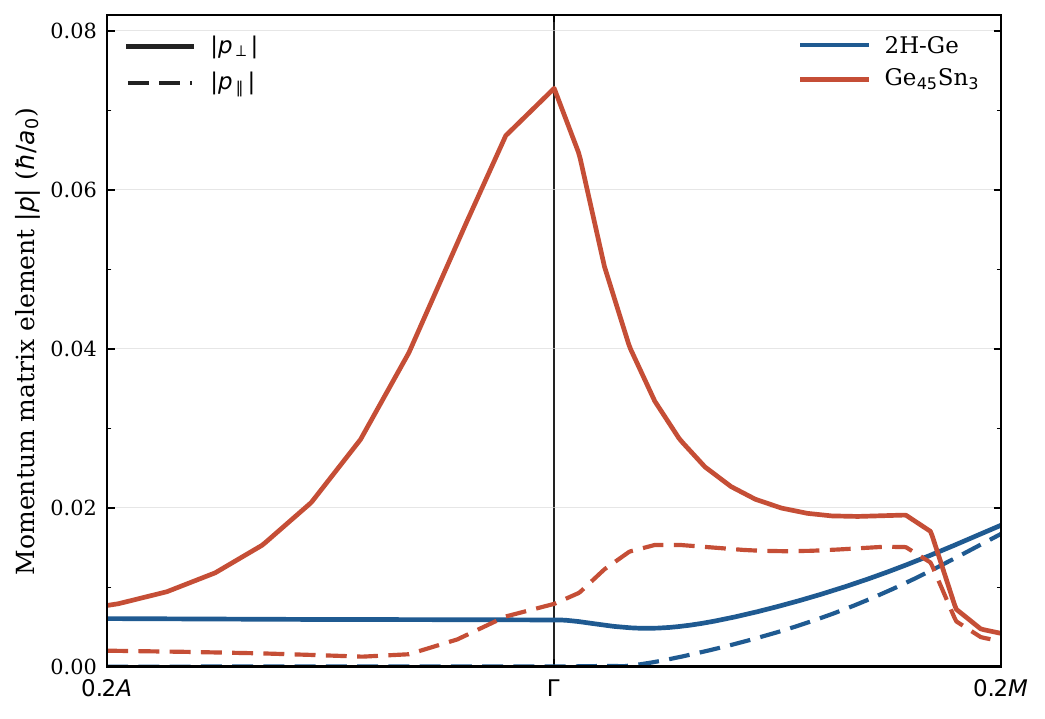}
      \caption{Momentum matrix elements for pristine 2H-Ge and a representative $\mathrm{Ge}_{45}\mathrm{Sn}_{3}$ along the $0.2A-\Gamma-0.2M$ path. Solid and dashed lines denote $|p_{\perp}|$ and $|p_{\parallel}|$, respectively. While the fundamental transition in pristine 2H-Ge remains nearly dark at $\Gamma$, Sn incorporation enhances $|p_{\perp}|$ by more than an order of magnitude and produces a finite parallel component through alloy-induced state mixing.}
    \label{fig:matrix_elements}
\end{figure}

To quantify the optical activity of the hexagonal alloys, we calculate the optical transition matrix elements $|p| = |\langle \psi_c | \mathbf{\hat{e}} \cdot \mathbf{p} | \psi_v \rangle|$ for the fundamental transition at the $\Gamma$ point, following the approach of Ref.~\cite{rodl2019accurate}. Due to the lonsdaleite lattice symmetry ($D_{6h}$), the optical response is highly anisotropic. The light-matter interaction strongly depends on the orientation of the polarization vector $\mathbf{\hat{e}}$ relative to the crystal $c$-axis.
As established by earlier first-principles calculations \cite{rodl2019accurate, Suckert2021} and subsequently analyzed using multiband $\mathbf{k} \cdot \mathbf{p}$ theory \cite{pulcu2024multiband}, the optical activity of the fundamental gap in pristine 2H-Ge is governed by symmetry selection rules.
 In the absence of spin-orbit coupling (SOC), the transition between the top valence band ($\Gamma_{9v}$) and the lowest conduction band ($\Gamma_{8c}$) is strictly dipole-forbidden. The introduction of SOC induces a hybridization of the $\Gamma_{8c}$ state with higher-lying conduction-band states. This spin-orbit mixing relaxes the selection rule, rendering the optical transition allowed while giving a large spatial anisotropy. Thus, the transition is dominant for light polarized perpendicular to the $c$-axis ($E \perp c$) and remains negligible for polarization parallel to the $c$-axis ($E \parallel c$). The corresponding polarization anisotropy in the vicinity of $\Gamma$ is illustrated in Fig.~\ref{fig:matrix_elements}
 
In pure 2H-Ge, this SOC-induced fundamental transition remains relatively weak, with $p_{\perp,\mathrm{CB}} = 5.893\times10^{-3} \hbar/a_0$ whereas the transition to CB+1 is strongly allowed, with $p_{\perp,\mathrm{CB+1}} = 0.394 \hbar/a_0$ as shown in Table~\ref{tab:momentum_matrix}. Hence, above the effectively  dark CB, pristine 2H-Ge contains a bright CB+1 carrying a greater optical weight.
The introduction of Sn into the lattice via random alloying lowers the local crystal field symmetry within the SQS supercell and  this microscopic symmetry breaking in principle relaxes the rigid $D_{6h}$ selection rules.
However, we find that for a given Sn concentration, the incorporation  of Sn atoms generally produces a configuration-dependent enhancement of the optical transition. 
This means that some SQS configurations retain an almost dark lowest CB, similar to 2H-Ge, whereas in others, the fundamental transition is enlarged by more than an order of magnitude.
A similar  observation was made  in the computational study of hexagonal SiGe alloys \cite{borlido2023ensemble}.
Therefore, we  averaged  the results of  different SQS's 
and report $p^{\rm eff}=\sqrt{\langle p^2\rangle}_{\mathrm{conf}}$ values in Table~\ref{tab:momentum_matrix} for each concentration. Although $p_{\perp,\mathrm{CB}}^2$ varies by more than two orders of magnitude across the alloy configurations, the combined strength $P_{\perp,\mathrm{pair}}^2=p_{\perp,\mathrm{CB}}^2+p_{\perp,\mathrm{CB+1}}^2$ remains within 0.112--0.139~$(\hbar/a_0)^2$. Thus, although the pair strength is not strictly conserved, Sn incorporation predominantly redistributes optical weight from CB+1 to CB.
\begin{table*}[t]
\caption{Effective momentum matrix element magnitudes for the transitions from the VB maximum to the two lowest CB states at $\Gamma$. All matrix elements are in units of $\hbar/a_0$. For the alloy configurations, $\left|p_{\alpha,n}\right|_{\mathrm{eff}}=\sqrt{\langle|p_{\alpha,n}|^2\rangle_{\mathrm{conf}}}$.}
\label{tab:momentum_matrix}
\begin{ruledtabular}
\begin{tabular}{lcccc}
Material & $\left|p_{\perp,\mathrm{CB}}\right|_{\mathrm{eff}}$ & $\left|p_{\parallel,\mathrm{CB}}\right|_{\mathrm{eff}}$ & $\left|p_{\perp,\mathrm{CB+1}}\right|_{\mathrm{eff}}$ & $\left|p_{\parallel,\mathrm{CB+1}}\right|_{\mathrm{eff}}$ \\
\hline
2H-Ge, mBJ-LDA (Ref.~\cite{rodl2019accurate}) & $5.920\times10^{-3}$ & $0$ & $3.940\times10^{-1}$ & $0$ \\
2H-Ge, this work & $5.893\times10^{-3}$ & 0 & $3.943\times10^{-1}$ & 0 \\
$\mathrm{Ge}_{47}\mathrm{Sn}_{1}$ & $4.005\times10^{-2}$ & $1.647\times10^{-3}$ & $3.710\times10^{-1}$ & $1.395\times10^{-2}$ \\
$\mathrm{Ge}_{46}\mathrm{Sn}_{2}$ & $5.361\times10^{-2}$ & $3.376\times10^{-3}$ & $3.533\times10^{-1}$ & $2.410\times10^{-2}$ \\
$\mathrm{Ge}_{45}\mathrm{Sn}_{3}$ & $6.321\times10^{-2}$ & $6.358\times10^{-3}$ & $3.465\times10^{-1}$ & $3.066\times10^{-2}$ \\
$\mathrm{Ge}_{44}\mathrm{Sn}_{4}$ & $7.449\times10^{-2}$ & $8.603\times10^{-3}$ & $3.402\times10^{-1}$ & $3.476\times10^{-2}$ \\
\end{tabular}
\end{ruledtabular}
\end{table*}

\subsection{Dielectric Function and Absorption Spectra}

The macroscopic linear optical response of the alloy is governed by the frequency-dependent complex dielectric tensor, $\varepsilon(\omega) = \varepsilon_1(\omega) + i\varepsilon_2(\omega)$. In the lonsdaleite phase, the underlying lattice symmetry imposes a strict uniaxial optical anisotropy, yielding distinct ordinary ($\varepsilon^{xx} = \varepsilon^{yy} \equiv \varepsilon^{\perp}$) and extraordinary ($\varepsilon^{zz} \equiv \varepsilon^{\parallel}$) tensor components. The real part at the zero-frequency limit, $\varepsilon_1(0)$, defines the static dielectric constant, a critical parameter that governs carrier screening and exciton binding energies. Pure 2H-Ge exhibits an intrinsically high static dielectric constant. Substitution of heavier, more polarizable Sn atoms into the 2H-Ge lattice monotonically increases $\varepsilon_1(0)$ while systematically redshifting the dominant interband transition peaks of the imaginary part, $\varepsilon_2(\omega)$, toward lower photon energies. Despite the local symmetry breaking introduced by the random alloy distribution, the global structural anisotropy remains dominant.

The practical utility of 2H-Ge$_{1-x}$Sn$_x$ for infrared optoelectronics is ultimately determined by its absorption coefficient $\alpha(\omega)$, which dictates the penetration depth and generation rate of optical carriers. This macroscopic absorption can be derived directly from the components of the complex dielectric tensor discussed above via the standard relation,
\begin{equation}
\alpha(\omega) = \frac{\sqrt{2}\omega}{c} \left[ \sqrt{\varepsilon_1(\omega)^2 + \varepsilon_2(\omega)^2} - \varepsilon_1(\omega) \right]^{\frac{1}{2}},
\label{eq:absorption}
\end{equation}
where $c$ is the speed of light in vacuum. Consistent with the unfolded effective band structures, the calculated absorption spectra exhibit a sharp, step-like onset characteristic of a direct-gap semiconductor. As the Sn concentration increases, this absorption edge redshifts, tracking the composition dependent narrowing of the $\Gamma$-valley minimum into the MIR, while the optical response remains strongly anisotropic.

\begin{figure}[t]
    \includegraphics[width=0.5\textwidth]{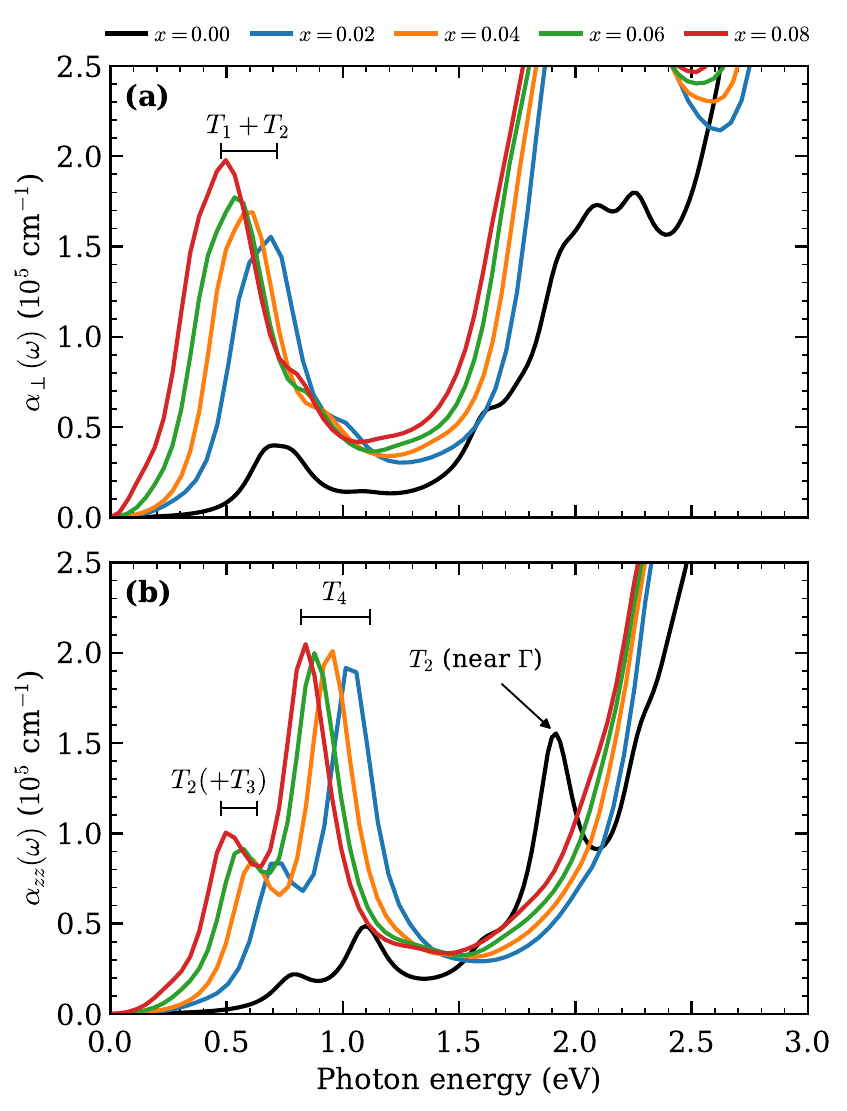}
            \caption{Calculated optical absorption spectra $\alpha(\omega)$ of 2H-Ge$_{1-x}$Sn$_x$ as a function of photon energy for (a) in-plane ($\alpha^{\perp} = (\alpha_{xx} + \alpha_{yy}) / 2$, $E \perp c$) and (b) out-of-plane ($\alpha^{zz}$, $E \parallel c$) polarization. The principal spectral features are labeled according to the transition assignments ($T_i$) defined and discussed in the text. Increasing Sn concentration redshifts the absorption edge and enhances the low-energy absorption.}
    \label{fig:absorption}
\end{figure}

Figure~\ref{fig:absorption} displays the computed absorption coefficients, derived via Eq.~\ref{eq:absorption} using VASPKIT \cite{wang2021vaspkit}, for the in-plane ($\alpha^{\perp} = (\alpha_{xx} + \alpha_{yy}) / 2$) and out-of-plane ($\alpha^{\parallel} \equiv \alpha_{zz}$) directions for pure 2H-Ge and four dilute Sn alloys. The alloy spectra are equal weight averages over five SQS realizations at each composition. The individual spectra and their full ranges are shown in Fig.~S3 of the Supplemental Material~\cite{supplementary}.
 Incorporation of Sn produces a pronounced redshift of the absorption edge and enhances the low-energy absorption in the near-IR/MIR range. Crucially, the macroscopic spectra inherit the polarization anisotropy of the transition matrix elements, with the in-plane response (\(E\perp c\)) dominating the band-edge absorption, while the out-of-plane response (\(E\parallel c\)) remains weak. To relate the absorption peaks appearing in Fig.~\ref{fig:absorption} to the band structures, we compared their spectral positions with the corresponding band energy separations and momentum matrix elements. Here, VB, $\text{VB}-1$, and $\text{VB}-2$ denote the three uppermost valence band Kramers pairs, while CB and $\text{CB}+1$ denote the two lowest conduction band pairs. The first peak in-plane contains overlapping contributions from $T_1$ ($\text{VB} \rightarrow \text{CB}+1$) and $T_2$ ($\text{VB}-1 \rightarrow \text{CB}+1$). In the out-of-plane spectrum, the lower maximum is dominated by $T_2$. Alloy disorder introduces an additional, smaller contribution from $T_3$ ($\text{VB}-2 \rightarrow \text{CB}$), which is suppressed in pristine 2H-Ge. The subsequent out-of-plane maximum is primarily associated with $T_4$ ($\text{VB}-2 \rightarrow \text{CB}+1$). These labellings remain unchanged across the investigated alloy compositions, while the corresponding transition energies shift systematically to lower energy with increasing Sn content.
The $E \parallel c$ peak near 1.9~eV in pure 2H-Ge is also primarily associated with the $\text{VB}-1 \rightarrow \text{CB}+1$ channel, but originates from finite in-plane wave vectors close to $\Gamma$, where the parallel polarized matrix element is strongly enhanced. However, the fundamental $\text{VB} \rightarrow \text{CB}$ transition determines the absorption edge but does not produce a resolved maximum. These trends indicate that 2H-Ge$_{1-x}$Sn$_x$ should exhibit strong uniaxial polarization anisotropy in emission and absorption, which may be useful for polarization-sensitive mid-IR devices.

\section{Conclusion}

In conclusion, we have carried out a first-principles investigation of the structural, electronic, and optical properties of dilute 2H-$\mathrm{Ge}_{1-x}\mathrm{Sn}_x$ alloys using SQS supercells, mBJ-LDA, and including spin-orbit coupling. The results show that the lonsdaleite lattice accommodates Sn substitution smoothly, with a monotonic expansion of the lattice parameters and only weak deviations from Vegard like behavior in the dilute regime. The nearly constant $c/a$ ratio shows that Sn incorporation produces an almost isotropic expansion of the average hexagonal lattice, without a pronounced macroscopic distortion along the $c$ axis. These structural trends provide the basis for the subsequent electronic and optical evolution of the alloy system.

As expected for standard alloy behavior in the dilute regime, the electronic structure exhibits a near-linear gap narrowing while retaining the band-edge character of the 2H-Ge host material. The mean gap decreases from 0.2992 eV in pristine 2H-Ge to 0.2521 eV at $x=0.0208$ and 0.1246 eV at $x=0.0833$, corresponding to a fitted reduction of approximately 20.4 meV per \% Sn. These alloy gaps correspond to wavelengths spanning approximately 5--10.0~$\mu\mathrm{m}$, demonstrating tunability across the MIR. The increasing configurational spread at higher Sn content further shows that the local Sn arrangement becomes important as the gap approaches zero.

Band unfolding confirms that the band-edge states retain a coherent Bloch character despite the alloy disorder, and the direct bandgap nature of the hexagonal host remains intact in the random alloy. The effective masses likewise preserve the strong anisotropy characteristic of the 2H-Ge, with their composition dependence is summarized in Table~\ref{tab:alloy_masses}. In particular, the in-plane conduction-band mass remains close to the 2H-Ge value over the compositions studied, whereas the out-of-plane mass varies more strongly with Sn content. The valence-band masses also change only moderately, indicating that the overall anisotropic band-edge character survives alloying.

The optical response follows the same symmetry-driven anisotropy. The disorder induced by the inclusion of Sn activates the weak fundamental VB-CB transition in a strongly configuration-dependent manner. The squared momentum matrix element strength of the two CB states indicates that alloying primarily redistributes optical weight between host derived dark (CB) and bright (CB+1) states. On average, the fundamental transition remains strongly enhanced for the polarization perpendicular to the $c$ axis, while transitions for $E \parallel c$ are weak compared to $E \perp c$. The trend is also reflected in the configurationally averaged absorption spectra, which show a polarization anisotropy, while the narrowing of the $\Gamma$-point gap shifts the absorption edge toward the MIR. Unlike for cubic GeSn, this tunabality does not require an indirect-to-direct crossover because 2H-Ge already possesses a direct $\Gamma$ point gap.

Comparison with hexagonal SiGe clarifies the role of Sn alloying. 2H-SiGe also exhibits symmetry driven anisotropic optical selection rules, but its usable spectral range is more limited because of the larger intrinsic gap of 2H-SiGe. In contrast, Sn incorporation into 2H-Ge drives a much stronger reduction of the fundamental gap, shifting the absorption edge efficiently into the mid-infrared already at low alloy concentrations. The similarity between the  polarization selection rules in 2H-SiGe and 2H-GeSn supports its origin in the band symmetry of the hexagonal host, whereas the smaller intrinsic gap of 2H-Ge makes Sn alloying particularly effective for mid-infrared tuning.

At higher Sn content, the bandgap becomes increasingly sensitive to the particular SQS realization and finite supercell size. Hence, the exact composition at which the gap closes cannot be determined unambiguously within the present work. Nevertheless, configurational ensembles establish gap narrowing throughout the dilute regime, while the unfolded band structures indicate that semiconducting configurations generally retain a $\Gamma$-centered band edge. Calculations employing larger supercells and many-body corrections would help to refine the quantitative gap-closure composition and optical spectra. Overall, the present results establish dilute 2H-$\mathrm{Ge}_{1-x}\mathrm{Sn}_x$ as a promising silicon-compatible material system, combining a direct $\Gamma$-point gap, alloy-enhanced band-edge optical activity, strong polarization anisotropy, and tunable mid-infrared response.
An important next step would be to explore the properties of the 4H and 6H  polytypes which have also been synthetized \cite{hex-GeSn-exp-jacs}.

\section{Acknowledgments}
We acknowledge financial support from the ONCHIPS project funded by the European Union’s Horizon Europe research and innovation programme under Grant Agreement No.~101080022. We also acknowledge the Digital Government Development and Project Management Ltd.\ for awarding us access to the Komondor HPC facility based in Hungary, as well as the Scientific Compute Cluster of the University of Konstanz
(SCCKN) for providing us access.

\bibliography{references}

\clearpage
\onecolumngrid
\pagestyle{empty}

\noindent\includegraphics[page=1,width=\textwidth]{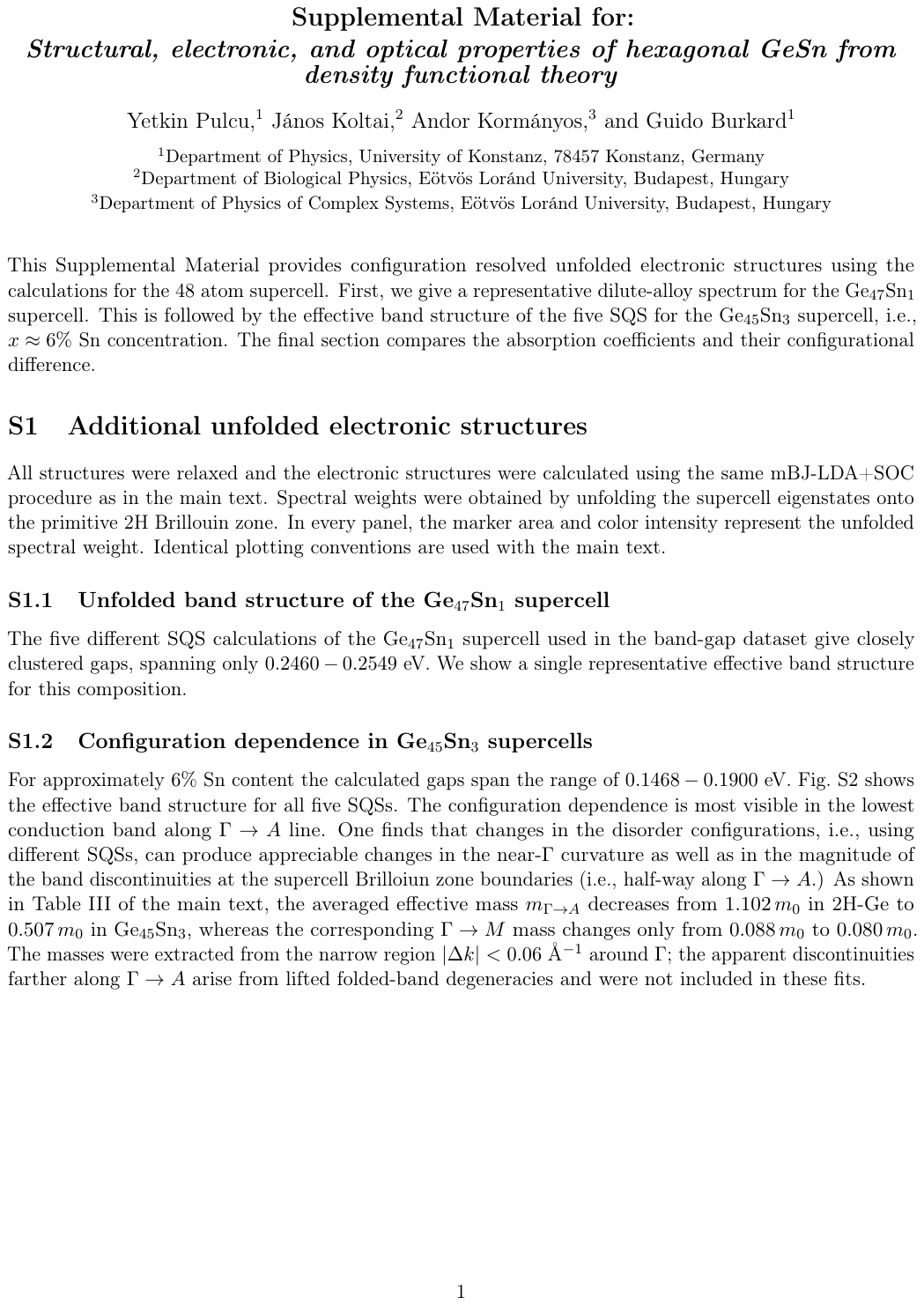}\clearpage
\noindent\includegraphics[page=2,width=\textwidth]{supplementary.pdf}\clearpage
\noindent\includegraphics[page=3,width=\textwidth]{supplementary.pdf}\clearpage
\noindent\includegraphics[page=4,width=\textwidth]{supplementary.pdf}\clearpage
\noindent\includegraphics[page=5,width=\textwidth]{supplementary.pdf}\clearpage

\end{document}